\documentclass[10pt,conference]{IEEEtran}
\IEEEoverridecommandlockouts
\usepackage{cite}
\usepackage{amsmath,amssymb,amsfonts}
\usepackage{algorithmic}
\usepackage{graphicx}
\usepackage{textcomp}
\usepackage{xcolor}
\def\BibTeX{{\rm B\kern-.05em{\sc i\kern-.025em b}\kern-.08em
    T\kern-.1667em\lower.7ex\hbox{E}\kern-.125emX}}

\usepackage[backgroundcolor=white,bordercolor=blue,linecolor=blue]{todonotes}
\usepackage{xspace}
\usepackage{blindtext}
\usepackage{hyperref}

\usepackage{multirow}
\usepackage{subcaption}
\usepackage{setspace}

\newcommand{\APPR}{\emph{MOTIF}\xspace}
\newcommand{\SEMU}{\emph{SEMu}\xspace}
\newcommand{\SEMUp}{\emph{SEMuP}\xspace}
\newcommand{\MOTIF}{\APPR}

\newcommand{\ASNLib}{\emph{ASN1lib}\xspace}

\newcommand{\UTIL}{\emph{LIBU}\xspace}

\newcommand{\MLFS}{\emph{MLFS}\xspace}

\usepackage{balance}
 
 \usepackage{amsmath}
\usepackage{array}
\usepackage{lipsum}
\usepackage{listings}
\definecolor{mGreen}{rgb}{0,0.6,0}
\definecolor{mGray}{rgb}{0.5,0.5,0.5}
\definecolor{mPurple}{rgb}{0.58,0,0.82}
\definecolor{backgroundColour}{rgb}{0.95,0.95,0.92}

\lstdefinestyle{CStyle}{
    backgroundcolor=\color{white},   
    commentstyle=\color{mGreen},
    keywordstyle=\color{magenta},
    numberstyle=\tiny\color{mGray},
    stringstyle=\color{mPurple},
    basicstyle=\scriptsize\ttfamily,
    breakatwhitespace=false,         
    breaklines=true,                 
    captionpos=b,                    
    keepspaces=true,                 
    numbers=left,                    
    numbersep=5pt,                  
    showspaces=false,                
    showstringspaces=false,
    showtabs=false,                  
    tabsize=1,
    language=C
}

\begin{document}

\title{Fuzzing for CPS Mutation Testing}

\author{\IEEEauthorblockN{Jaekwon Lee$^{\star,\ddagger}$, Enrico Viganò$^{\star}$, Oscar Cornejo$^{\star}$, Fabrizio Pastore$^{\star}$, Lionel Briand$^{\star,\ddagger}$}
\IEEEauthorblockA{\textit{$^{\star}$University of Luxembourg}, $^{\ddagger}$\textit{University of Ottawa} \\
\textit{Luxembourg, LU; Ottawa, CA}\\\{jaekwon.lee,enrico.vigano,oscar.cornejo,fabrizio.pastore,lionel.briand\}@uni.lu}
}

\maketitle

\begin{abstract}
Mutation testing can help reduce the risks of releasing faulty software. For such reason, it is a desired practice for the development of embedded software running in safety-critical cyber-physical systems (CPS). 
Unfortunately, state-of-the-art test data generation techniques for mutation testing of C and C++ software, two typical languages for CPS software, rely on symbolic execution, whose limitations often prevent its application (e.g., it cannot test black-box components).

We propose a mutation testing approach that leverages fuzz testing, which has proved effective with C and C++ software. Fuzz testing automatically generates diverse test inputs that exercise program branches in a varied number of ways and, therefore, exercise statements in different program states, thus maximizing the likelihood of killing mutants, our objective.

We performed an empirical assessment of our approach with software components used in satellite systems currently in orbit. Our empirical evaluation shows that mutation testing based on fuzz testing kills a significantly higher proportion of live mutants than symbolic execution (i.e., up to an additional 47
percentage points). Further, when symbolic execution cannot be applied, fuzz testing provides significant benefits (i.e., up to 41\% mutants killed). Our study is the first one comparing fuzz testing and symbolic execution for mutation testing; our results provide guidance towards the development of fuzz testing tools dedicated to mutation testing.
\end{abstract}


\section{Introduction}
\label{sec:introduction}

Software testing plays a key role in verifying and validating embedded software for cyber-physical systems (CPS). Ensuring the high-quality of test suites is therefore essential for quality assurance purposes in such contexts.

Mutation analysis is an effective approach to assess the quality of a test suite. Indeed, it entails measuring the mutation score, which is the proportion of programs with artificially injected faults (i.e., mutants) detected by the test suite~\cite{papadakis2019mutation}, and 
there exists a strong association between a high mutation score and a high fault revealing capability for test suites~\cite{papadakis2018mutation,Chekam:17}.
Further, recent work has shown that mutation analysis can be cost-effectively applied to large CPS software by combining multiple optimizations~\cite{Oscar:TSE}.

In practice, mutation analysis warrants the effective selection of inputs for mutation testing since test cases are required to detect all or a large proportion of the generated mutants.
A mutant detected by a test suite is said to be killed.
However, due to the typically high number of mutants generated in the context of large CPS projects~\cite{Oscar:TSE}, it is challenging for engineers to perform mutation testing manually. 

Unfortunately, we lack automated test data generation techniques (\emph{automated mutation testing} techniques) applicable to CPS; indeed, most of the existing techniques do not target the C and C++ languages, which are widely used in CPS domains.

The state-of-the-art (SOTA) solution for the automated mutation testing of C software (i.e., \SEMU~\cite{Chekam2021}) is based on the KLEE symbolic execution engine~\cite{KLEE}. Though it has shown to be effective with command line utilities, it inherits the limitations of symbolic execution. Specifically, it requires modeling of the environment (e.g., network communication) and cannot deal with programs that require complex analyses to enable input generation (e.g., programs with floating point instructions). Further, it generates test inputs for command line utilities, which are seldom used in CPS, and does not generate unit test cases nor target other CPS interfaces. Search-based techniques developed for other programming languages (e.g., Java)~\cite{fraser2011mutation} are impractical for C and C++ software because of the difficulty of instrumenting the software to compute dedicated fitness functions (e.g., branch distance).
For example, to compute branch distance at runtime, it is necessary to modify all the conditional statements in the software under test (SUT) and, for that, the source code must be processed with static analysis tools that require loading all the dependencies. Unfortunately, for large systems, this often leads to configuring such tools to process several source files in nested directories, which is impractical except if the tool is well integrated  with the compiler already in use for the SUT. Moreover, CPS source files often rely on architecture-specific C constructs (e.g., for the RTEMS compiler~\cite{RTEMS}) that are not successfully parsed by static analysis frameworks~\cite{Oscar:TSE}.

In this paper, we propose relying on gray-box fuzz testing techniques~\cite{manes2019art}, also called fuzzing techniques or fuzzers, to generate test data for mutation testing.
Grey-box fuzzers apply evolutionary algorithms to data (called seed data), which is either randomly generated or user-provided, to generate test input data (usually input files) that maximize code coverage and trigger failures. In contrast to other search-based testing techniques~\cite{EvoSuite}, they do not rely on branch distance~\cite{stvr.294} but instead make use of coarser heuristics like bucket-based branch coverage, which analyzes if an input exercises a branch for a number of times not observed before (see Section~\ref{sec:background:fuzzing}); such heuristics can be implemented with simple extensions of standard C/C++ compilers, which facilitates their adoption in industry.

Because of their effectiveness and applicability to C and C++ software, gray-box fuzzers are promising alternatives to support the automated mutation testing of CPS software. However, fuzzers target console software, while CPS software is usually tested either with system-level test scripts interacting with a hardware emulator or through unit and integration test cases implemented with the same language as the SUT. In this paper, we focus on the automated generation of unit test cases because fuzzing large systems is an open research problem~\cite{KimXSO22}. For simplicity, we use the term unit test cases to indicate test cases implemented with the same language as the SUT providing inputs to a function under test; however, the unit test cases generated by our approach may exercise either single units (e.g., a C function) or multiple components (e.g., if the function under test invokes other functions or interacts with remote components through the network). 

Since fuzzers cannot automatically generate test drivers for unit testing, several techniques that generate test drivers to enable fuzz testing of C and C++ APIs have been developed~\cite{FuzzGen,FUDGE,APICRAFT}; however, they are not maintained and therefore not applicable in practice. Further, they require the availability of client programs using the API under test, which are not available in our context.

To address the limitations above, as a first contribution of this paper, we propose an approach based on fuzzing that consists of an automated pipeline supporting the generation of unit test cases for mutation testing; we call our approach \MOTIF (MutatiOn TestIng with Fuzzing). Our pipeline includes the automated generation of seed data and the automated generation of test drivers; to enable mutation testing, test drivers automatically determine if the outputs of the mutant differ from the outputs of the original software.
We do not design a dedicated fuzzing algorithm but rather propose an approach to apply SOTA fuzzers to support automated mutation testing. Our intuition is that the bucket-based coverage strategy of fuzzers can effectively drive the selection of test inputs that kill mutants because such strategy, in addition to selecting inputs leading to different software states, can also track, and be guided by, differences in the behavior of the original and the mutated function.
From a practical standpoint, relying on standard fuzzing algorithms helps with the adoption of our solution by practitioners because the maintenance of well-known fuzzing tools is guaranteed by several interest groups (e.g., companies investing in reliability and security).

As a second contribution, we compare \MOTIF with a SOTA symbolic execution approach to determine if fuzzing is more effective and can overcome the limitations of symbolic execution in practical settings. For our experiments, we considered three software components used in satellites currently in orbit: \MLFS, a mathematical library qualified by the European Space Agency (ESA) for flight systems, \UTIL{}, a utility library for nanosatellites developed by one of our industry partners in a project with ESA~\cite{FAQAS}, and \ASNLib, a serialization/deserialization library generated with the ESA ASN.1 compiler~\cite{mamais:hal-02263447}.

Our empirical results show that, in the two case study subjects where symbolic execution is applicable (\ASNLib and part of \UTIL{}), \MOTIF outperforms mutation testing based on symbolic execution by 46.86 and 10.52 percentage points, respectively. For subjects in which symbolic execution is not applicable (\MLFS and part of \UTIL{}), \MOTIF achieves interesting results by killing 35.97\% and 41.38\% of the live mutants, respectively.

This paper proceeds as follows. Section~\ref{sec:related} provides related work on automated mutation testing and background on symbolic execution and fuzzing. Section~\ref{sec:approach} describes our pipeline for automated mutation testing with fuzzing and an alternative pipeline relying on symbolic execution. Section~\ref{sec:evaluation} presents our empirical evaluation. Section~\ref{sec:conclusion} concludes the paper.

\section{Background and Related Work}
\label{sec:related}

This paper relates to techniques for automated mutation testing and fuzzing; relevant work is discussed below.

\subsection{Symbolic execution}
\label{sec:symbex}

Symbolic execution (SE) is a program analysis technique that relies on an interpreter to process the source code of the SUT and automatically generate test inputs~\cite{Anand2013}. 
Inputs are represented through symbolic values; during the symbolic execution,  the state of the SUT includes the symbolic values of program variables at that execution point, a path constraint on the symbolic values to reach that point, and a program counter. 
The path constraint is a boolean formula that captures the conditions that the inputs must satisfy to follow that path. Constraint solving~\cite{SATandCPsurvey:2006} is then used to identify assignments for the symbolic inputs that  satisfy the path constraint.

SE presents several limitations, including (1) the need for abstract representations  for the external environment and any black-box components used by the SUT (otherwise, the SE engine cannot know what outputs to expect from the environment), (2) path explosion (the SE engine may need to process a large number of paths before satisfying a target predicate), (3) path divergence (i.e., abstract representations do not behave like the real systems), (4) handling of complex constraints (e.g., solving constraints with floating point variables).

A recent solution to partially address the above-mentioned limitations is 
 dynamic symbolic execution (DSE), which consists of treating only a portion of the program state symbolically. Concrete program states help dealing with complex constraints or path explosion (e.g., SE is used after a certain branch has already been reached using a concrete input).
 However, most frameworks with DSE capabilities like Angr~\cite{shoshitaishvili2016state}, KLEE~\cite{KLEE}, and S2E~\cite{S2E} rely on binary interpretation, which, in practice, requires some degree of environment modeling (e.g., libc library modeling in KLEE) and limit their practical applicability~\cite{SymCC}.

Compilation-based approaches like QSYM~\cite{yun2018qsym} and SYMCC~\cite{SymCC} augment the original program with instructions to populate and solve symbolic expressions while the original software is executed; such characteristic eliminates some limitations of interpretation-based approaches  thus being applicable to a broader set of software systems. For example, since the symbolic execution interacts with the actual environment there is no need to emulate it within the interpretation layer. SYMCC requires the source code of the SUT, while QSYM relies on dynamic binary instrumentation. However, we have excluded QSYM and SYMCC from our investigation because there is no dedicated mutation testing approach for them. Further, implementing such a mutation testing approach  is a significant research challenge since it entails finding solutions to integrate, within the original program, the logic to derive inputs that kill mutants. Last, these approaches have shown to provide their best results when combined with fuzzing, a solution referred to as \emph{hybrid fuzzing} (see Section~\ref{sec:background:fuzzing}). \emph{Hybrid fuzzing} is nevertheless difficult to apply in our context because QSYM and SYMCC still present technical limitations preventing their application to CPS software; indeed, SYMCC relies on LLVM, which is not applicable to certain systems~\cite{Oscar:TSE,DAMAT}, while QSYM is not maintained~\cite{QSYM:repository} and relies on an outdated version of the PIN instrumenter~\cite{PIN:paper,PIN:tool}.

\subsection{Fuzzing}
\label{sec:background:fuzzing}

Fuzzing (or fuzz testing) is an automated testing technique 
that generates test inputs by repeatedly modifying\footnote{To avoid confusion, we avoid the term `mutation' when describing fuzzing techniques.} existing inputs; the selection of the inputs to modify is usually driven by metrics collected during the execution of the SUT.
Based on the information collected during program execution, fuzzing techniques (i.e., fuzzers) are classified as black-box, white-box, or gray-box. 

In this paper, we focus on \emph{grey-box fuzzers} because they have demonstrated to effectively maximize code coverage~\cite{FuzzBenchPaper} and discover faults~\cite{ICST:22:Sarro} (mainly crashes and memory errors), two objectives that relate to the problem studied in this paper; indeed, to kill a mutant it is necessary to (1) exercise a mutated statement, which can be achieved by maximizing code coverage, and (2) exercise the mutated statement with many different inputs (i.e., in different states), a common practice in fuzzers to discover crashes and memory errors.

Most fuzzers generate input files to be used for system-level testing of console applications; however, engineers can implement driver programs (hereafter, \emph{fuzzing drivers}) that rely on the data generated by the fuzzer to test other software interfaces (e.g., APIs, see Section~\ref{sec:motif:step1}). Most fuzzers keep a pool of input files and rely on the following evolutionary search process: (1) select an input file from the pool, (2) modify the input file to generate new input files, (3) provide the new input files to the SUT and monitor its execution, (4) report crashes or problems detected through sanitizers~\cite{UBSan}, (5) add to the pool all the input files that contribute to improve code coverage.

What facilitates the adoption of fuzzers is that they rely on simple dynamic analysis strategies to trace branch coverage of C/C++ programs. A common strategy consists of dynamically identifying branches by applying a hashing function to the identifiers assigned to code blocks by compile-time instrumentation; it is implemented as an extension of popular C/C++ compilers~\cite{AFL:instr}. Further, instead of relying on traditional branch coverage~\cite{ammann2016introduction}, most fuzzers adopt a bucketing approach to track the number of times each branch has been covered by each input file across ranges: only once, twice, three times, between four and seven, between 8 and 15, between 16 and 31, etc.; the fuzzers add to the pool those files that cover at least one branch for a range of times (i.e., a bucket) not observed before. Such bucketing strategy help reach software states that are not reachable by simply relying on branch coverage. 

Fuzzers mainly differ with respect to the strategy adopted to (1) select what operations to apply in order to modify input files and obtain new ones (e.g., MOpt~\cite{MOPT} relies on a particle swarm optimization algorithm) and (2) select the inputs from the input pool (e.g., AFLfast~\cite{AFLfast} and AFL++~\cite{AFL++} rely on a simulated annealing algorithm and prioritize new paths and paths exercised less frequently). Also, fuzzers differ in the strategy adopted to determine interesting inputs. For example, \emph{directed grey-box fuzzers}~\cite{WangDGF2020}, instead of maximizing code coverage, aim to reach specific targets --- usually a subset of program locations (e.g., modified code) or invalid sequences of operations (e.g., use-after-free). 
 \emph{Hybrid fuzzers}~\cite{Hybrid,Stephens2016,yun2018qsym}, instead, rely on grey-box fuzzing to explore most of the execution paths of a program and leverage DSE to explore branches that are guarded by narrow-ranged constraints when the fuzzer does not improve coverage further. 
 The two SOTA hybrid fuzzers combine AFL~\cite{AFL} with QSYM~\cite{yun2018qsym} and SYMCC~\cite{SymCC}.

Some researchers have addressed the problem of generating test drivers to fuzz test program functions as in unit testing~\cite{FuzzGen,FUDGE,APICRAFT}; however, they all target library APIs and make the assumption that such APIs have been already integrated into consumer programs (i.e., programs using the library API).
FuzzGen~\cite{FuzzGen} relies on the static analysis of both the API under test and its consumers to derive call graphs capturing valid sequences of function invocations and derive test drivers. 
Different from FuzzGen, 
Fudge~\cite{FUDGE} works with a single API consumer and, instead of synthesizing test drivers from a graph, it relies on code snippets (i.e., sequences of API calls and the variables in scope) extracted from the consumer. 
ApiCraft~\cite{APICRAFT} targets APIs without source code and leverages both static and dynamic information (headers, binaries, and traces) to collect control and data dependencies for API functions.
Unfortunately, consumer programs are not available when performing mutation analysis for CPS, which makes the above-mentioned approaches inapplicable; indeed, some components are often developed only for a specific product (e.g., the application layer for a satellite under development), while other components, despite being implemented for reuse (e.g., utility libraries), should be verified by mutation testing before they are integrated into consumer programs.

Other techniques address the problem of generating highly structured input files~\cite{TensileFuzz,Skyfire}. TensileFuzz generates structured inputs (e.g., image or zip files) by probing random executions to derive constraints for potential input fields, and then relying on string constraint solving to derive inputs~\cite{TensileFuzz}. SkyFire, instead, learns a probabilistic context-sensitive grammar to generate JSON and XML files~\cite{Skyfire}. Such techniques can generate input files with a complex structure but they do not generate unit test cases, which is necessary in our context; however, leveraging those approaches to  populate complex data structures may also help with unit-level fuzz testing.

\subsection{Automated mutation testing}

To kill a mutant, a test case should satisfy three conditions: \emph{reachability} (i.e., the test case should execute the mutated statement), \emph{necessity} (i.e., the test case should cause an incorrect intermediate state if it reaches the mutated statement), and \emph{sufficiency} (i.e., the observable state of the mutated program should differ from that of the original program)~\cite{offutt1997automatically}. 
Automated mutation testing approaches differ regarding the strategy adopted to satisfy these conditions.

There exist two families of 
automated mutation testing techniques based respectively on:
\emph{constraint solving} and \emph{meta-heuristic search}.
Only one of them relies on fuzzing~\cite{brown2020mutation}, as further described below.

In this Section, we mainly focus on techniques targeting C and C++ programs because these languages are used in many CPS; unfortunately, the C and C++ languages are more complex to process for static and dynamic analysis techniques than the higher-level languages targeted by most of the techniques in the literature (e.g., Java).

\subsubsection{Techniques based on constraint solving}
\label{sec:back:cs}

Inspired by the earlier work of Offut et al.~\cite{offutt1997automatically}, Holling et al. execute symbolically the original and mutated functions with input data leading them to generate different outputs~\cite{holling2016nequivack}. A similar technique from Riener et al.~\cite{riener2011test} relies on a bounded model checker (BMC) to select the input values that kill the mutant. 
Unfortunately, no prototype tools for the above-mentioned approaches are available.

The SOTA tool for automated mutation testing is \SEMU~\cite{Chekam2021,SEMUgit}, which relies on KLEE to generate test inputs based on SE. 
To speed up mutation testing, \SEMU relies on meta-mutants (i.e., it compiles mutated statements and the original statements together). First, \SEMU relies on SE to reach mutated statements (reachability condition).
Then, for each mutant, it relies on constraint solving to determine if inputs that weakly kill the mutant exist (necessity condition). For killable mutants, it symbolically runs the mutated and the original program in parallel; when an output statement is reached (e.g., a \texttt{printf} or the \texttt{return} statements of the main function), it relies on constraint solving to identify input values that satisfy the sufficiency condition.

\subsubsection{Techniques based on meta-heuristic search}

Most of the work on automated mutation testing with meta-heuristic search targets Java software; we report the most relevant techniques below.
Ayari et al.~\cite{ayari2007automatic} rely on an Ant Colony Optimization algorithm~\cite{dorigo2006ant}
 driven by a fitness function that focuses on the reachability condition. Precisely, their fitness measures the distance (number of basic blocks in the program's control flow graph) between the mutated statement and the closest statement reached by a test case. 
Fraser and Zeller~\cite{fraser2011mutation}, instead, extended the EvoSuite tool~\cite{EvoSuite} with a fitness function considering the reachability and the necessity conditions (number of statements that are covered a different number of times by the original and the mutated program). The integration of mutation testing into EvoSuite has been further improved with branch distance metrics tailored to the operator used to generate the mutants~\cite{fraser2015achieving}. Recently, EvoSuite has been further extended by Almulla et al. with adaptive fitness function selection (AFFS), a hyperheuristic approach that relies on reinforcement learning (RL) algorithms to determine which composition of fitness functions to use~\cite{Almulla2022}. 
Unfortunately, when applied to mutation testing, AFFS does not perform better than SOTA solutions~\cite{fraser2015achieving}.

Concerning C software, we should note the work of Souza et al.~\cite{souza2016strong}, who rely on the Hill Climbing AVM algorithm~\cite{KorelAVM}.
They combine three fitness functions that rely on branch distance to measure how far an input is from satisfying each of the three killing conditions.
The mutation score obtained with simple C programs ranges between 52\% and 93\%.
The approach has been implemented on top of AUSTIN, a search-based test generation tool for C~\cite{lakhotia2010austin,LAKHOTIA2013112,AUSTIN}; however, this implementation is not available.
A recent search-based testing tool prototype for C is Ocelot~\cite{Scalabrino:18}; however, it has not been extended for automated mutation testing. 
Another key limitation of both Ocelot and AUSTIN is that they implement preprocessing steps that do not work with complex program structures (e.g., we couldn't apply them to the subject programs considered in our empirical evaluation because of preprocessing errors).

A recent mutation testing technique targeting C software is that of Dang et al.~\cite{dang2019efficiently}, who propose a co-evolutionary algorithm that reduces the search domain at each iteration (the original search domain is replaced by the joint domain of the best solutions found); unfortunately, their prototype is not available.

\subsection{Techniques based on fuzzing}
The work of Bingham~\cite{brown2020mutation} is the only one to rely on fuzzing to automate mutation testing for C software.
For input generation, it relies on TOFU~\cite{wang2020tofu}, a grey-box, grammar-aware
fuzzer that
generates grammar-valid inputs by modifying existing ones. 
Similar to Ayari's work, TOFU's input generation strategy is guided by the distance between the mutated statement and the closest statement reached by a test case; however, instead of generating unit test cases, it generates input files matching a given grammar.
Unfortunately, the results obtained by Bingham are preliminary (they targeted only the Space benchmark~\cite{SPACE}) and a prototype tool is not available.

Mu2~\cite{MU2}, which has been developed in parallel with \MOTIF, is a fuzzer that integrates the findings of search-based unit test generation~\cite{EvoSuiteMutation} to generate test input files with fuzzing: it relies on the mutation score to drive the generation of test inputs. Different from \MOTIF, Mu2 tests every live mutant with each generated input and, in the file pool, prioritizes those files that increase the mutation score; the scalability of such choice is enabled by dynamic classloading and instrumentation, two options that are feasible for Java programs but not for the C/C++ programs targeted by \MOTIF. Further, by targeting Java, Mu2 can easily determine if mutants are killed by relying on the method `equals', which is implemented by every class to determine if two instances are equal; the method `equals' is not available in C and C++ software. Results show that Mu2 kills more mutants than the inputs generated by a traditional fuzzer; however, the question remains if Mu2 (i.e., testing all the live mutants together) is more effective than the approach used by \MOTIF (i.e., testing the original and mutated function in sequence).
Mu2's results follow previous work showing that, in Java benchmarks, prioritizing inputs that increase the mutation score may lead to higher branch coverage and mutation score than traditional prioritization strategies based on branch coverage~\cite{CoverageGuided}.

To summarize, our research is motivated by the lack of support for automated mutation testing of C/C++ software. The SOTA approach for the automated mutation testing of C/C++ software (i.e., \SEMU) relies on KLEE and inherits its limitations, making it inapplicable to most CPS software; further, it does not generate unit test cases but selects inputs for console programs. Other SE tools (QSYM and SYMCC) also present technical limitations preventing their application to CPS software. Search-based approaches for the mutation testing of C/C++ software present acute feasibility challenges because of the difficulty of executing static analysis, which is needed for branch distance fitness, in large software projects. Though fuzzing appears to be a feasible input generation strategy for mutation testing, existing fuzzers do not generate test drivers for unit testing. The only fuzzer proposed for mutation testing is not available for download and its results are very preliminary.

\section{Proposed Approaches}
\label{sec:approach}

In this Section, we present two approaches for automated mutation testing: (1) MutatiOn Testing wIth Fuzzing (\APPR), our main contribution, which is automated through a pipeline of commands to generate unit test cases by relying on fuzzing (Section~\ref{sec:motif}). (2) \SEMUp, which is a pipeline derived from \MOTIF to perform unit mutation testing with \SEMU (Section~\ref{sec:semu}). They are both used in our empirical evaluation.

\subsection{MOTIF}

\label{sec:motif}
\begin{figure}[tb]
\begin{center}
\includegraphics[width=8.4cm]{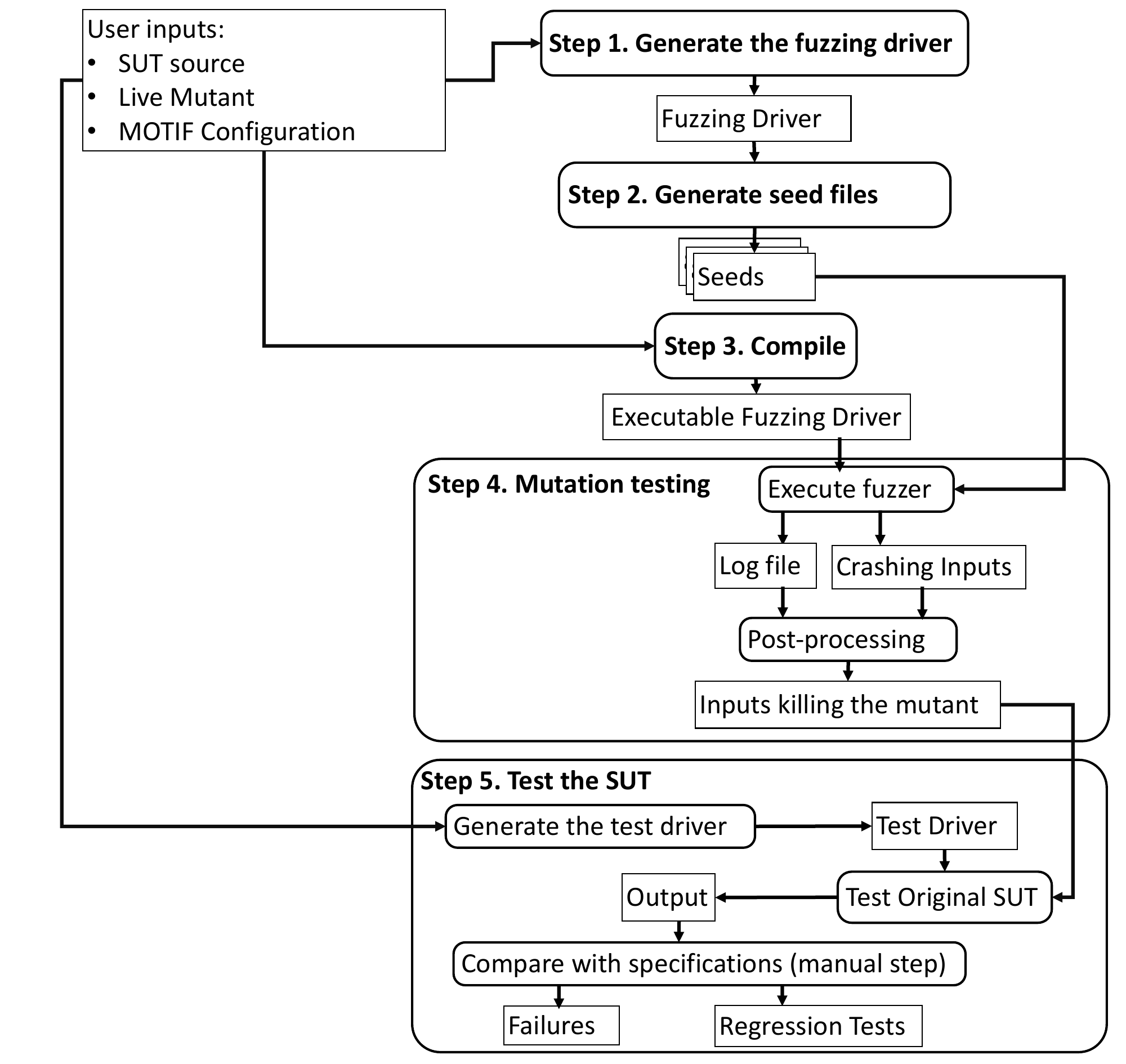}
\caption{The \MOTIF process.}
\label{fig:motif:process}
\end{center}
\end{figure}

Similar to Holling et al.~\cite{holling2016nequivack}, we aim to identify a set of test inputs that lead to different outputs when provided to the original and to the mutated function. To achieve such objective with fuzzing, for each mutated function, \APPR generates a fuzzing driver that reads the input data generated by the fuzzer and then appropriately provides such data, as arguments, to both the original and the mutated function. Finally, the fuzzing driver compares the output data generated by the original and the mutated function, if they differ, the mutant has been killed. 

Our intuition is that fuzzers not only help kill mutants because they can achieve high coverage~\cite{FuzzBenchPaper} and reach multiple program states, including faulty ones~\cite{ICST:22:Sarro}, but also that, by invoking the original and the mutated functions within a same fuzzing driver, we can leverage the bucket-based fuzzing strategy to drive the generation of test inputs towards killing mutants. 
Indeed, if differences in the coverage of the original and the mutated function are observed then the two functions behave differently and, consequently, they yield different outputs leading to the mutant being killed~\cite{schuler2013covering,schuler2009efficient,Oscar:TSE}.
Also, large differences in coverage lead to new buckets being covered and since fuzzing favors inputs covering new buckets, it indirectly leads to inputs killing mutants. 
In other words, the bucket-based fuzzing strategy may help kill mutants by preserving, during test generation, those inputs that lead to incorrect intermediate states but do not kill the mutant (i.e., they do not satisfy the sufficiency condition); the following iterations of the evolutionary search process implemented by the fuzzer (see Section~\ref{sec:background:fuzzing}) may modify such inputs such that, in addition to reaching an incorrect intermediate state, they also satisfy the sufficiency condition.
We leave to future work the extension of fuzzers with dedicated strategies; for example, instead of measuring the coverage of the mutated function, the fuzzer may measure the difference in coverage between  the original and the mutated function, and use this information to further prioritize the inputs in the fuzzer queue (e.g., test first the inputs that leads to larger differences).

\APPR automatically generates all the scaffolding required to test the original function, the mutated function, and compare their outputs. Specifically, \APPR implements the workflow depicted in Figure~\ref{fig:motif:process}, which consists of five Steps that we describe below. 

\APPR receives as input a set of mutants (source files) to be killed; each mutant matches the original source file except for the statements modified by a mutation operator. 
The \APPR Steps in Figure~\ref{fig:motif:process} are repeated for each mutant. However, Steps 1, 2, 3, and part of Step 5 (i.e., \emph{Generate the test driver}) can be executed only once for all the mutants belonging to a same function; indeed, the structure of the input and output data of a function does not change based on the mutants---we do not target interface mutation~\cite{delamaro2001interface}.

\subsubsection{Step 1 -- Generate the fuzzing driver}
\label{sec:motif:step1}
In Step 1, MOTIF relies on the \emph{clang} static analysis library~\cite{CLANG} to process the SUT and determine the types of the parameters required by the function under test. Such information is used to generate a fuzzing driver for mutation testing; an example fuzzing driver for the function \texttt{T\_POS\_IsConstraintValid} belonging to our \ASNLib{} case study subject is shown in Listing~\ref{asn_driver}. The fuzzing driver renames the mutated function by adding the prefix \emph{mut\_}.

\begin{lstlisting}[style=CStyle, tabsize=2, caption=Example fuzzing driver for the \ASNLib{} subject., label=asn_driver, mathescape=true]
int main(int argc, char** argv){
  load_file(argv[1]);  // load the input file and
  // extends the input with random data if needed

  /* Variables for the original function */
  T_POS origin_pVal;   // for the first parameter
  int origin_pErrCode; // for the second parameter
  /* Variables for the mutated function */
  T_POS mut_pVal;      // for the first parameter
  int mut_pErrCode;    // for the second parameter
  /* Variables for the return values */
  flag origin_return;  // for the original 
  flag mut_return;     // for the mutant

  /* Copy the input data to the variables for the original function */
  get_value(&origin_pVal, sizeof(origin_pVal), 0); 
  get_value(&origin_pErrCode,sizeof(origin_pErrCode),0); 
  log("Calling the original function");
  origin_return = T_POS_IsConstraintValid(&origin_pVal, &origin_pErrCode);

  /* Copy the same input data to the variables for the mutated function */
  seek_data_index(0); //reset the input data pointer
  get_value(&mut_pVal, sizeof(mut_pVal), 0); 
  get_value(&mut_pErrCode, sizeof(mut_pErrCode), 0); 
  log("Calling the mutated function");
  mut_return = mut_T_POS_IsConstraintValid(&mut_pVal, &mut_pErrCode);

  log("Comparing result values: ");
  ret += compare_value(&origin_pVal, &mut_pVal, sizeof(origin_pVal));
  ret += compare_value(&origin_pErrCode,&mut_pErrCode, sizeof(origin_pErrCode));
  ret += compare_value(&origin_return, &mut_return, sizeof(origin_return));

  if (ret != 0){
    log("Mutant killed");
    safe_abort();
  }
  log("Mutant alive");
  return 0;
}    
\end{lstlisting}

The fuzzing driver contains two sets of variables (Lines 5-7 and 8-10) whose types match the parameters of the function under test and are provided as input to the original and to the mutated function, respectively; in our example, it declares a \texttt{struct T\_POS} and an \texttt{int} variable. 
The two sets of variables are then assigned by performing a byte-by-byte copy of a same portion of  input file provided by the fuzzer (Lines 16-17 and 23-24); \APPR ensures to copy a number of bytes to match the size of the assigned variable. If the input file provided by the fuzzer is shorter than required, \APPR extends it with random data (Line~2). Additionally, the fuzzing driver declares the variables required to store the functions' return values (Lines 11 to 13).

The original and the mutated functions are then invoked (Lines 19 and 26).
The fuzzing driver then compares the output generated by the original and the mutated function (Line 28-31). Unfortunately, in C and C++, the presence of pointer and reference arguments complicates distinguishing  input and output parameters. Further, to determine input parameters, we cannot rely on data-flow analysis because it entails preprocessing the SUT with a static analysis framework (e.g., LLVM~\cite{LLVM}), which is often not feasible with CPS software~\cite{Oscar:TSE}. Therefore, we adopt a simple solution consisting of comparing all the parameters and return values of the original and mutated function; indeed, 
comparing input parameters does not lead to incorrect mutant killing since they are not modified.
For pointers, we compare the pointed data (e.g., an \texttt{int} instance for \texttt{int*}). If the pointer is used as an array, the end-user can specify the expected length of the array, so the array data can be compared. 
When arrays are inputs to the function under test, the end-user may not need to provide the length because \MOTIF automatically generates arrays with a default length (100). 
If arrays are dynamically allocated by the function under test, the end-user should specify the minimal possible length (e.g., an array of length one), to avoid false positives due to readings out of the array bounds.
For data structures with pointer fields, it is possible to specify the pointed data length and the initialization procedure.
When the outputs differ, the fuzzing driver stops its execution with an abort signal (Line 35 in Listing~\ref{asn_driver}) thus letting the fuzzer detect the aborted execution and store the input file; \MOTIF then stops the fuzzer because the mutant has been killed.

\begin{table}[t]
\centering
\scriptsize
\caption{Seeds assigned to types}
\label{tab:seedValues}
\begin{tabular}{
|@{\hspace{1pt}}>{\raggedleft\arraybackslash}p{10mm}@{\hspace{1pt}}|
@{\hspace{1pt}}>{\raggedleft\arraybackslash}p{27mm}@{\hspace{1pt}}|
@{\hspace{1pt}}>{\raggedleft\arraybackslash}p{20mm}@{\hspace{1pt}}|
@{\hspace{1pt}}>{\raggedleft\arraybackslash}p{27mm}@{\hspace{1pt}}|}
 \hline
 \textbf{Type}& \textbf{Seed 1}&\textbf{Seed 2}&\textbf{Seed 3}\\
 \hline
int& -1& 0& 1\\
Bool& False& True& \\
float& -3230283776.0& 0.0& 1072693248.0\\ double& 13826050856027422720.0& 0.0&
4602891378046628864.0\\
char& 0xFF& 0x00& 0x41\\
byte& 0xFF& 0x00& 0x41\\
ISO8601& 2145916800.999999999&
1970-01-01T00:00:00Z& 2038-01-01T00:00:00Z\\ 
\hline
\end{tabular}%

\end{table}

\subsubsection{Step 2 -- Generate seed files}

In Step 2, \APPR generates seed files based on the types of input parameters for the function under test. Seed files are used by the fuzzer to start the testing process; usually, fuzzers are executed with seed files that correspond to typical inputs for the SUT. In our case, we automatically generate seed files that contain enough bytes to fill all the input parameters with values covering basic cases. Precisely, for each primitive type, we have identified three seed values that are representative of typical input partitions; they are reported in Table~\ref{tab:seedValues}. For example, for numeric values, we provide zero, a negative, and a positive number. Based on these seed values, for each fuzzing driver, \APPR generates at most three seed files in such a way  that each seed value is covered 
at least once for every input parameter. %

\begin{figure}[tb]
\begin{center}
\includegraphics[width=8.4cm]{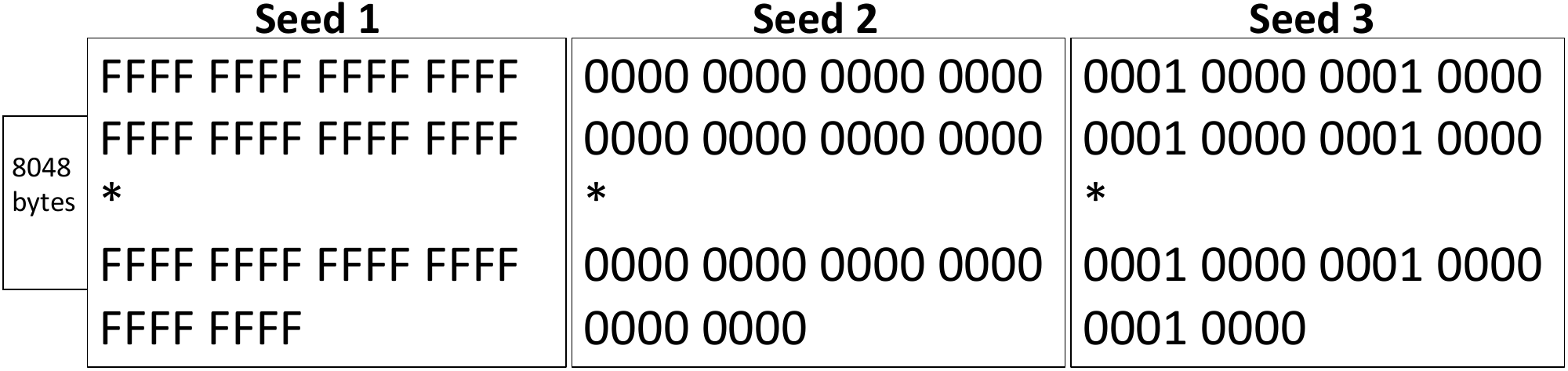}
\caption{Seed files generated for the fuzzing driver in Listing~\ref{asn_driver}.}
\label{fig:seedFiles}
\end{center}
\end{figure}

Example seed files for function \texttt{T\_POS\-\_IsConstr\-aint\-Valid} are provided in Figure~\ref{fig:seedFiles} (type definitions in Listing~\ref{asn_types}).
Please note that \MOTIF can generate seed files also for complex input types, 
indeed the \texttt{struct T\_POS} received as input by 
function \texttt{T\_POS\_IsConstraintValid} consists of an \texttt{enum} (named \emph{kind}), which is used to specify the type of data stored inside the rest of the struct, and a union (named \emph{u}), which is sufficiently large to contain the data for all the data types selectable with the variable \emph{kind}. \MOTIF treats such struct as an \texttt{int} array thus filling it with the seeds \emph{0xFFFFFFFF}, \emph{0x00000000}, and \emph{0x00000001}. 
The first four bytes in the seed files (see Figure~\ref{fig:seedFiles}) belong to the \texttt{enum} item \emph{kind}, and are filled with the seed values of the \texttt{int} type. The same happens for the \texttt{union} field \emph{u} but, since the \texttt{union} has a size of 8,052 bytes (size of \emph{subTypeArray} with 4 bytes padding\footnote{https://research.nccgroup.com/2019/10/30/padding-the-struct-how-a-compiler-optimization-can-disclose-stack-memory/}), \MOTIF repeats the same set of four bytes 2,013 times. The last four bytes belong to the second parameter of \texttt{T\_POS\_IsConstraintValid}, the \texttt{int} \emph{*\_pErrCode}.

\begin{lstlisting}[
style=CStyle, 
tabsize=2,
caption=Definition of \texttt{struct T\_POS}, 
label=asn_types, 
mathescape=true
]
typedef enum { T_POS_NONE,           longitude_PRESENT, 
               latitude_PRESENT,     height_PRESENT, 
               subTypeArray_PRESENT, label_PRESENT, 
               intArray_PRESENT,     myIntSet_PRESENT, 
               myIntSetOf_PRESENT,   anInt_PRESENT
} T_POS_selection;

typedef struct {
    T_POS_selection kind;
    union { asn1Real longitude; asn1Real latitude; 
            asn1Real height;    My2ndInt anInt; 
            T_POS_label label;  T_ARR intArray; 
            T_SET myIntSet;     T_SETOF myIntSetOf; 
            T_POS_subTypeArray subTypeArray; 
    } u;
} T_POS;
\end{lstlisting}

\subsubsection{Step 3 -- Compile the SUT}

In Step 3, \APPR compiles the fuzzing driver, the mutated function, and the SUT using the fuzzer compiler; this is necessary to collect the code coverage information required by the fuzzer.

\subsubsection{Step 4 -- Perform mutation testing}
\label{sec:stepFour}
In Step 4, \APPR runs the fuzzer to generate inputs for the executable fuzzing driver. The fuzzer keeps generating input files until it reports one or more crashes, after which MOTIF stops the fuzzer. The execution leads to the generation of fuzzing driver logs and crashing inputs (i.e., input files that caused a crash during the execution of the fuzzing driver). Since fuzzers generate several input files from each input taken from the file pool, and since all of them are executed by the fuzzer, more than one crashing input may be reported.

Fuzzing driver logs include checkpoints indicating the progress of testing (see Lines 18, 25, 28, 34, 37 in Listing~\ref{asn_driver}).
For each crashing input, \APPR processes the corresponding logs to distinguish between:
\begin{itemize}
\item Crashes occurring during the execution of the original function. They indicate either the presence of a fault in the original function or the violation of preconditions. We ignore these inputs because they do not correspond to inputs killing a mutant.
\item Crashes occurring during the execution of the mutated function. Since the crashes occur during the execution of the mutated function, which is executed after the original one, we can safely conclude that the test inputs do not cause any crash in the original function. Therefore, the observed crashes indicate that the mutant introduced a fault that was exercised by the input. Thus, we can consider these inputs as inputs that kill the mutant.
\item Aborted executions due to the fuzzing driver determining that the mutant has been killed (see Line 35 in Listing~\ref{asn_driver}).
\end{itemize}

\MOTIF keeps all the test inputs killing a mutant. 
However, the function under test may generate non-deterministic outputs and in such situations, despite observed differences in outputs, 
the inputs may not have killed the mutant. 
 For example, two consecutive invocations of a  function that reads and writes global variables may lead to different outputs even if the mutated statement is not exercised; consequently, the input suggested by the fuzzer would be a false positive.  
To minimize false positives, \MOTIF automatically re-executes every test input killing a mutant with a modified version of the fuzzing driver that
invokes the original function instead of the mutated function. If this false positive driver, as we refer to it, reports a difference in the outputs of the two function calls, it implies that the function under test is non-deterministic and thus that the input does not kill the mutant. \MOTIF considers mutants exclusively killed by false positive inputs to be live. 
 To kill mutants in functions modifying global state variables, the end-user should manually introduce the instructions required to reset the state between the two function calls in the fuzzing driver, which is similar to what required by  other fuzzing approaches for unit and library testing (e.g., LibFuzzer~\cite{LibFuzzer}).

\subsubsection{Step 5 -- Test the SUT}

In this Step, \APPR generates a test driver for the SUT. An example test driver for function \texttt{T\_POS\_IsCon\-straintValid} is shown in Listing~\ref{asn_test}. The test driver matches the fuzzing driver except that (1) it invokes only the original function (Line 5 in Listing~\ref{asn_test}) and (2) instead of comparing the outputs obtained from two function invocations, it prints out the output data generated by the original function (Lines 7 to 9). The test driver is used to test the original SUT with the inputs that kill the mutant and the outputs should then be verified by a software engineer based on the SUT specifications. If the observed output values are correct, they can be used as oracles for future regression testing. Otherwise, a fault has been found in the SUT; such a scenario is one of the key advantages of mutation testing: by testing the SUT with inputs that detect simulated human mistakes (mutants), actual faults in the SUT are more likely to be found than with randomly selected inputs.

In our test driver, the print statements for struct and pointers are generated based on the configuration of the fuzzing drivers. Precisely, by default, all the bytes belonging to a struct are printed out. In the presence of pointers, if the end-user specifies the size of the data referred to by pointers, the test driver prints the pointed data instead of the pointer value.

\begin{lstlisting}[style=CStyle, tabsize=2, caption=Example test driver for the \ASNLib{} subject., label=asn_test, mathescape=true]
int main(int argc, char** argv){
  load_file(argv[1]); /* load the input file */
  // Declaration of variables and assignment with input file data missing to save space...
  /* Invoke the original function*/
  _return = T_POS_IsConstraintValid(&pVal, &pErrCode);
  /* Print output values of the original function */
  printf_struct("pVal (T_POS)=", &pVal, sizeof(pVal));
  printf("pErrCode (int) = %
  printf("return (flag) = %
  return 0;
}
\end{lstlisting}

\subsection{\SEMUp}
\label{sec:semu}

To compare the effectiveness of fuzzing and SE when used for automated mutation testing, we have adapted the \MOTIF pipeline to enable test generation with \SEMU; we call the adapted pipeline \SEMUp. At a high level, \SEMUp follows the same steps of \MOTIF, with differences concerning how  inputs and outputs are declared to enable test generation with SE.

\begin{lstlisting}[style=CStyle, tabsize=2, caption=Example \SEMU driver corresponding to the fuzzing driver in~Listing~\ref{asn_driver}., label=semu_driver, mathescape=true]
int main(int argc, char** argv){
    // Declare variable to hold function returned value
    _Bool result;
    // Declare arguments and make input ones symbolic
    T_POS pVal;
    int pErrCode;
    
    klee_make_symbolic(&pVal, sizeof(pVal), "pVal");
    // Call function under test
    result = T_POS_IsConstraintValid(&pVal, &pErrCode);
    // Print output data
    printf("pErrCode = %
    printf("result = %
    return (int)result;
}
    
\end{lstlisting}

In Step 1, we generate \emph{SEMu drivers} instead of fuzzing drivers; an example \SEMU driver generated for function \texttt{T\_POS\_IsCon\-straint\-Valid} is shown in Listing~\ref{semu_driver}. In \SEMU drivers it is necessary to specify what are the input parameters to be treated symbolically (see Line 8 in Listing~\ref{semu_driver}); input parameters are provided as configuration parameters by the end-user. \SEMU drivers do not include explicit comparisons between the outputs of the mutated and the original function because such comparison is taken care by \SEMU when symbolically executing the original and the mutated functions in parallel (see Section~\ref{sec:back:cs}). Precisely, the \SEMU driver invokes only the function under test and prints to standard output the data values that should be considered to determine if a mutant has been killed. Similar to \MOTIF, \SEMU also requires end-users to manually specify how to process data values belonging to data structures referenced with pointers.  

For \SEMU, there is no Step 2 (i.e., we do not generate seed inputs). In Step 3, we compile the mutated function and the \SEMU drivers with LLVM. Step 4 concerns the execution of \SEMU and the processing of its logs to determine if mutants have been killed. Step 5 is conceptually the same as for \MOTIF, except that we load the inputs generated by KLEE.

\section{Empirical Evaluation}
\label{sec:evaluation}

We address the following research questions:

\emph{RQ1. How does mutation testing based on fuzzing compare to mutation testing based on symbolic execution, for software where the latter is applicable?}
SE proved to be an effective mean to perform mutation testing of command line tools that do not rely on floating-point instructions nor integrate black-box components. Therefore, SE may still outperform fuzzing when applied to mutation testing of CPS units that satisfy those assumptions.

\emph{RQ2. How does mutation testing based on fuzzing perform with  software that cannot be tested with symbolic execution?} The motivation for our work stems from the limited applicability of SE and we therefore aim to assess if fuzzing can effectively overcome such limitations.

\emph{RQ3. How does \MOTIF's seeding strategy contribute to its results?}
\MOTIF kills mutants either through the generated seeds or through the inputs generated by the fuzzer; we therefore aim to assess how the two strategies individually contribute to \MOTIF results.

\subsection{Subjects of the study}

\begin{table}[tb]
\caption{Subject artifacts.}
\label{table:caseStudies} 
\footnotesize
\centering
\begin{tabular}{|
@{\hspace{1pt}}p{9mm}
@{\hspace{2pt}}|
@{\hspace{1pt}}>{\raggedleft\arraybackslash}p{16mm}@{\hspace{1pt}}|
@{\hspace{1pt}}>{\raggedleft\arraybackslash}p{8mm}@{\hspace{1pt}}|
@{\hspace{1pt}}>{\raggedleft\arraybackslash}p{15mm}@{\hspace{1pt}}|
@{\hspace{1pt}}>{\raggedleft\arraybackslash}p{26mm}@{\hspace{1pt}}|
@{\hspace{1pt}}>{\raggedleft\arraybackslash}p{9mm}@{\hspace{1pt}}|
}
\hline
\textbf{Subject}&\textbf{Open-source}&\textbf{LOC}&\textbf{\# Test cases}&\textbf{Statements} \textbf{coverage}&\textbf{MS}\\
\hline
\MLFS{}{}& Yes& 5,402 &  4,042 & 100.00\% & 81.80\%\\
\UTIL{}& No& 10,576 & 201 & 83.20\% & 71.20\%\\
\ASNLib{}{}& Yes& 7,260 &  139 & 95.80\% &58.31\% \\
\hline
\end{tabular}

\end{table}

To address our research questions, we considered software deployed on space CPS (satellites) currently in orbit.
This included (a) \MLFS{}, the Mathematical Library for Flight Software~\cite{MLFS}, which complies with the ECSS criticality category B~\cite{ecss40C,ecss80C}, (b) \UTIL{}, which is a utility library developed by one of our industry partners and used in NanoSatellites, 
and (c) \ASNLib{}, a serialization/deserialization library generated with ASN1SCC from a test grammar provided by ESA. ASN1SCC is a compiler that generates C/C++ code suitable for low resource environments~\cite{ASN1CC,ASN1:paper}.

Our software subjects are provided with test suites; information about their code coverage is reported in Table~\ref{table:caseStudies}.
Some of our test suites do not achieve 100\% statement coverage because some components need to be tested with specific hardware not available to us; therefore, we generated mutants only for the  covered statements.
We generated mutants with MASS~\cite{Oscar:TSE,MASSTOOL}; specifically, we rely on all the mutation operators supported by MASS and which proved effective in previous experiments on similar subjects. We excluded mutants that are identified as equivalent or duplicate according to trivial compiler equivalence methods~\cite{Oscar:TSE}. Column \emph{MS} in Table~\ref{table:caseStudies} provides the mutation score for our case study subjects; it corresponds to the proportion of mutants detected by the test suite. The highest mutation score is observed with MLFS, whose test suite achieves MC/DC adequacy~\cite{chilenski1994applicability}. The lowest mutation score is observed with \ASNLib{}, which is automatically generated by the ASN1SCC using a grammar-based approach~\cite{mamais:hal-02263447}. Our subjects' mutation score is in line with empirical investigations reporting mutation scores ranging from 55\% to  95\%~\cite{Ramler2017,delgado2018evaluation}, for CPS software.

To perform test data generation, we rely on the mutants not killed by the original test suites. We assume that the live mutants are not equivalent (i.e., produce the same outputs for every input) to the original software and though this could be an under-approximation, it does not introduce bias in the comparison between \MOTIF and \SEMUp, which both cannot kill equivalent mutants. Further, two mutants $m_a$ and $m_b$ can also be duplicates (i.e., they lead to the same outputs for every input) or subsumed (i.e., $m_a$ is killed by a superset of the test cases killing $m_b$). 
However, the identification of test inputs that kill mutants is a precondition to determine if mutants are duplicate or subsumed~\cite{Shin:TSE:DCriterion:2018}; for this reason, including duplicate and subsumed mutants should not introduce bias in the comparison of the two approaches. In other words, a mutation testing approach should easily kill mutants that are either duplicates and subsume other killed mutants; if it does not happen, it is correct to penalize such an approach in the empirical evaluation.
Finally, for \UTIL{}, we have excluded 8 mutants manually identified as equivalent after inspecting the (few) live mutants not killed by \MOTIF for RQ1; we could not perform the same analysis for the other cases as such manual analysis would take too long.

\subsection{Experimental setup}
\label{sec:experiment:setup}

We performed our experiments using a prototype implementation of the \MOTIF and \SEMUp pipelines described in Section~\ref{sec:approach}.

For \MOTIF, as fuzzer, we selected AFL++ because it is the fuzzer that performed better in terms of code coverage, according to a recent benchmark in the literature~\cite{FuzzBenchPaper}; moreover, along with HonggFuzz~\cite{honggfuzz}, it is the fuzzer that maximizes fault coverage in another recent benchmark~\cite{ICST:22:Sarro}.

Since the number of live mutants is large for complex CPS, we assume that an effective setup for mutation testing consists in relying on distributed services that enable the execution of a large number of computing nodes in parallel; for example, we execute our experiments on a grid infrastructure. Although multiple mutants may be killed by similar test inputs~\cite{Chekam2021}, we do not test live mutants with inputs that have killed other mutants because we test mutants in parallel. In the future, we will assess how MOTIF's effectiveness can be improved by reusing inputs that have killed mutants.

To account for randomness factors in \MOTIF  and \SEMUp, we executed each approach ten times for each subject. For each mutant, we executed both \MOTIF and \SEMUp for 10,000 seconds, which we determined, in a preliminary study, to be sufficient for \SEMUp to maximize the percentage of killed mutants. Precisely, for \SEMUp we allocate 10,000 seconds to the symbolic execution process, which means that, after the timeout, if the mutant has not been killed yet, \SEMUp still tries to generate test inputs using the path conditions traversed so far, which leads to an execution time for \SEMUp that is slightly higher than \MOTIF's (around 650 seconds more).

\MOTIF is available online~\cite{MOTIFGIT}; also, we provide a \emph{replication package} with our open-source subjects and all our empirical data~\cite{REPLICABILITY}.

\begin{figure*}[tb!]
	\begin{subfigure}[t]{0.49\textwidth}\centering
	    \includegraphics[width=\columnwidth]{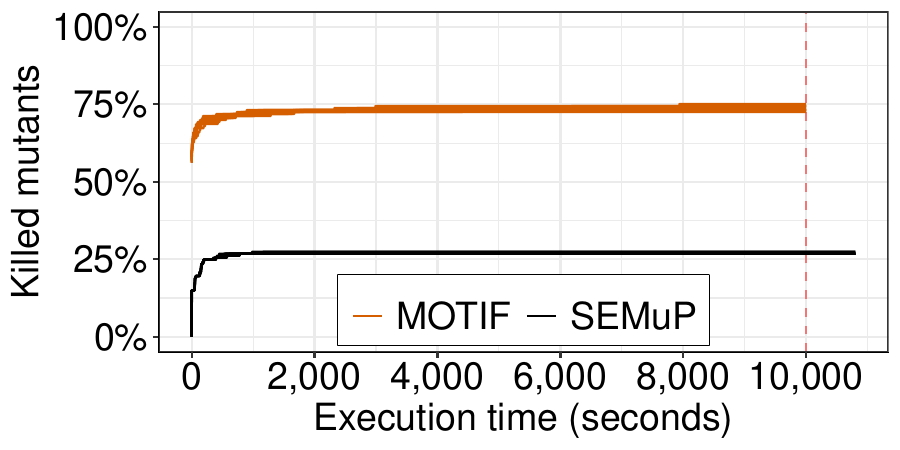}
		\caption{\UTIL{}}
	    \label{fig:rq1 libutil}
	\end{subfigure}
	\hspace{0em}
	\begin{subfigure}[t]{0.49\textwidth}\centering
	    \includegraphics[width=\columnwidth]{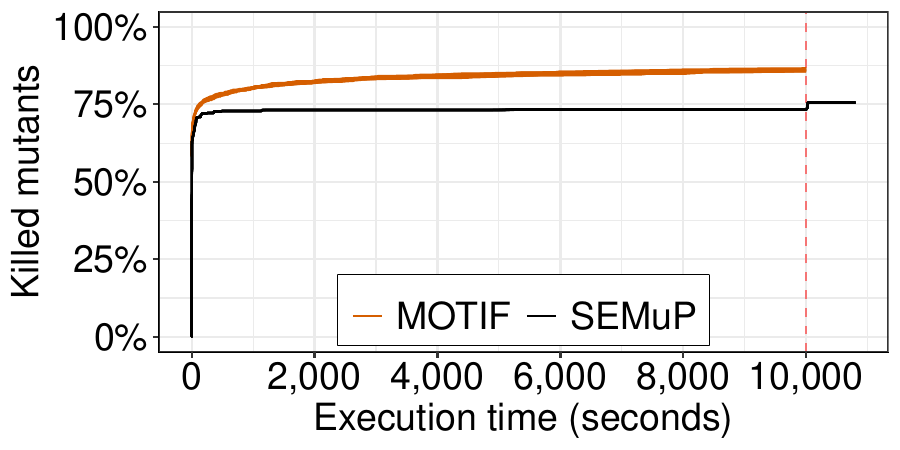}
		\caption{\ASNLib{}}
	    \label{fig:rq1 asn}
	\end{subfigure}
\caption{Percentage of live mutants killed by \MOTIF and \SEMUp}
\label{fig:rq1}
\end{figure*}

\subsection{RQ1 - Fuzzing vs Symbolic Execution}

\subsubsection{Design}

We compare fuzzing and symbolic execution in terms of cost-effectiveness.
The effectiveness of an automated mutation testing tool can be measured in terms of the proportion of live mutants killed. Its cost depends on the time required to kill the mutants; indeed, lengthy test data generation may delay the testing process and increase the usage of computing resources.
Cost is also driven by the time required to manually inspect test outputs; 
however, \MOTIF and \SEMUp should require the same manual inspection time because they invoke the same functions under test and print out the same output values.
Therefore, regarding cost, we focus on execution time and thus compare cost-effectiveness in terms of live mutants killed for different time budgets. 

To address RQ1, we could not consider \MLFS because it works mainly with floating point arguments, which are not supported by KLEE. An old version of KLEE addresses floating point variables but it is not integrated into \SEMU. We therefore focus on \UTIL{} and \ASNLib{}; however, for \UTIL{} we considered only four out of 27 source files, because all the other source files included I/O operations, which are not supported by KLEE/\SEMU, or cannot be compiled into LLVM bitcode. This leads to 1,347 live mutants for \ASNLib{} and 153 for \UTIL{}. 

\subsubsection{Results}

Figure~\ref{fig:rq1} depicts the percentage of live mutants killed by \MOTIF and \SEMUp for \UTIL{} (\ref{fig:rq1 libutil}) and \ASNLib{} (\ref{fig:rq1 asn}), respectively.
Each plot depicts the percentage of mutants killed after each second, for each run. Separate curves are plotted to visualize dispersion across the ten runs.
 The vertical dashed line shows the 10,000 seconds timeout when, for \ASNLib{}, which includes paths with  several nested  conditions, we observe a rapid increase in the number of mutants killed by \SEMUp. At that point, \SEMUp stops exploring paths and generates inputs that satisfy the current path condition, which, sometimes, is sufficient to identify inputs that kill mutants. 

The plots show that \MOTIF outperforms \SEMUp. After 10,000 seconds, \MOTIF kills between 111 (72.55\%) and 115 (75.16\%) mutants for \UTIL{} (avg. is 112.9, 73.79\%) and between 1,153 (85.60\%) and 1,167 (86.64\%) for \ASNLib{} (avg. is 1,159.5, 86.08\%). In contrast, \SEMUp kills 41 (26.80\%) to 42 (27.45\%) mutants for \UTIL{} (avg. is 41.2, 26.93\%) and 1,017 (75.50\%) to 1,018 (75.58\%) for \ASNLib{} (avg. is 1,017.8, 75.56\%). 
On average, across the ten runs, \MOTIF kills a percentage of mutants that is 46.86 
percentage points (pp) and 10.52 pp 
higher than \SEMUp's, for \UTIL{} and \ASNLib{}, respectively.

The difference between \MOTIF and \SEMUp is significant at every timestamp, based on Fisher test
\footnote{We compare the proportion of mutants killed by the two approaches across the ten experiments, which gives us high-statistical power given the large number of mutants.}~\cite{Fisher} 
($\alpha < 0.01$). For example, after one minute, \MOTIF kills, on average, 101.3 (\UTIL{}) and 976.2 (\ASNLib{}) mutants, while \SEMUp kills
29 (\UTIL{}) and 924.6 (\ASNLib{}) mutants. 
For \UTIL{}, \MOTIF quickly reaches a near plateau because of \UTIL{}'s simple control logic.

Though \MOTIF outperforms \SEMUp, they show some degree of  complementarity, which suggests that future work should integrate hybrid fuzzers in \MOTIF (see Section~\ref{sec:background:fuzzing}, including the limited applicability of existing hybrid fuzzers). If we consider the best run of each approach, in the case of \ASNLib{}, \MOTIF kills 252 (18.70\%) mutants not killed by \SEMUp, while \SEMUp kills 103 (7.65\%) mutants not killed by \MOTIF. 
In the case of \UTIL{}, \MOTIF kills 74 (48.36\%) mutants not killed by \SEMUp, while \SEMUp kills 1 (0.65\%) mutant not killed by \MOTIF. 
We manually inspected some of the mutants and noticed that \SEMUp is sometimes better at generating inputs that satisfy narrow, simple constraints. However, such a characteristic is more useful for \ASNLib{}, which mainly performs boundary checks for nested data structures, rather than the utility library. On the other hand, \MOTIF is better when \SEMUp fails to solve complex constraints. For example, for \UTIL{}, \SEMUp could not kill 52 mutants affecting a conditional statement with 24 bitwise operations, 44 mutants affecting a conditional statement with 13 conditions expressed using inequalities, and 5 mutants affecting the size of the buffer used in \texttt{snprintf} statements. Finally, \MOTIF enabled the discovery of four bugs in \UTIL{} that were confirmed by developers; \SEMUp discovered three of them too.

\subsection{RQ2 - Fuzzing effectiveness}

\subsubsection{Design}

The mutants considered for RQ2 are the ones that cannot be tested with \SEMUp because 
of limitations of symbolic execution, which appear to be prevalent in the context of CPS. Recall that such software often cannot be compiled with LLVM, include I/O operations, and rely on floating point variables. To determine if fuzzing is effective at compensating for the limitations of symbolic execution, we applied \MOTIF to all the mutants of \MLFS (3,891) and a subset of the mutants derived from the \UTIL{} functions excluded for RQ1 (290). Precisely, for \UTIL{}, we selected all the mutants for which we can derive a complete fuzzing driver automatically (i.e., eight functions that can be tested without executing setup operations, 94 mutants) and a random subset of the other functions (32 functions, 196 mutants).

As in RQ1, we discuss cost-effectiveness to determine if fuzzing can effectively overcome the limitations of symbolic execution.

\subsubsection{Results}

\begin{figure}[tb]
\begin{center}
\includegraphics[width=8.4cm]{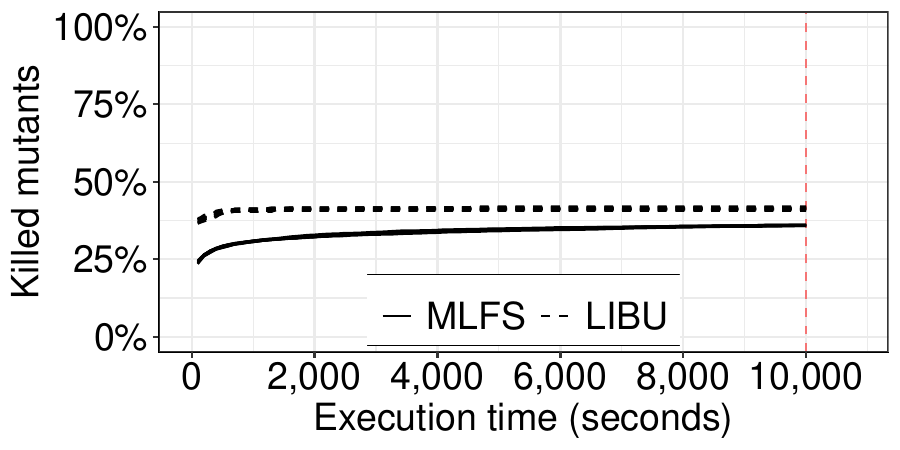}
\caption{RQ2 results with \MLFS{} and \UTIL{}.}
\vspace*{-1em}
\label{fig:results:mlfs}
\end{center}
\end{figure}

Figure~\ref{fig:results:mlfs} shows the percentage of live mutants killed by \MOTIF for \MLFS and \UTIL{}. 

In the case of \MLFS, 
after 10,000 seconds, MOTIF kills on average 1,399.4 (35.97\%) mutants (min 1,391, max 1,408).
The proportion of killed mutants is lower than for RQ1 because of, most probably, the mathematical nature of \MLFS, resulting in mutants being killed only by inputs from a very small part of the input domain.

For \UTIL{}, after 10,000 seconds, MOTIF kills, on average, 120 (41.38\%) mutants (min 118, max 122). Such percentage of killed mutants is again lower than for RQ1 because some of the 
mutants can be killed only with inputs belonging to a narrow portion of a large input domain (e.g., an input string that matches a string stored in a global variable).

As for RQ1, for both \MLFS and \UTIL{}, after one minute, \MOTIF kills a large proportion of the mutants killed after 10,000 seconds. On average, after one minute, \MOTIF kills 63.39\% (i.e., $22.80\%/35.97\%$) of all the  mutants killed for \MLFS and 88.26\% ($36.52\%/41.38\%$) of all the \UTIL{} mutants killed. Our results show that \MOTIF can be practically useful even when the budget available for mutation testing is limited. 

For \UTIL{}, the number of mutants killed by \MOTIF reaches a plateau after 1,500 seconds (25 minutes). For \MLFS, the number of killed mutants keeps increasing over time, thus suggesting that a large test budget may help \MOTIF identify inputs that kill mutants when they belong to a narrow subdomain of the input space.

\subsection{RQ3 - Seeding effectiveness}

\subsubsection{Design}

To discuss how \MOTIF seeds  contribute to mutation testing results, we focus on the proportion of mutants killed with seed inputs in the experiments performed to address RQ1 and RQ2. 

\subsubsection{Results}

In RQ1 experiments, for \UTIL{} and \ASNLib{}, one mutant (less than 1\% of the mutants killed on average in 10,000 seconds) and 280 (24.15\%) mutants are killed by seeds, respectively. In RQ2 experiments, 
seeds kill 76 \MLFS mutants (5.43\%) and 26 \UTIL{} mutants (21.66\%).

The percentage of mutants killed by seed inputs largely depends on the nature of the functions under test. For \MLFS and the \UTIL{} functions considered for RQ1, such percentage is low because they mainly alter mathematical operations whose mutants are killed with inputs satisfying complex constraints. For RQ2, the proportion of \UTIL{} mutants killed is higher because several mutants 
alter conditions verifying the correctness of input strings; the seed strings generated by \MOTIF include characters (e.g., spaces) that are targeted by such correctness controls, thus killing the mutants. 
Seed inputs do not introduce bias in RQ1 results since \SEMUp kills most of the mutants killed by seed inputs ($267/280$ for \ASNLib{} and $1/1$ for \UTIL{}).

Concluding, although the selected seed inputs help kill mutants, the contribution of the fuzzing process is significant with, at the very minimum (RQ2-\ASNLib{}), 75.85\% (i.e., $100\%-24.15\%$) of the killed mutants being killed by fuzzing.

\subsection{Threats to validity}

To address threats to internal validity, we manually verified that \MOTIF and \SEMUp correctly execute and, further, we manually inspected a large subset of the generated test cases and all the mutants killed by \MOTIF but not \SEMUp. Further, our false positive driver ensures that \MOTIF results are not affected by the presence of global variables or, more generally, non-determinism. Although we do not reset global state variables in fuzzing drivers, note that across all experiment runs, out of 27,918 mutants reported as killed by the fuzzing driver, only 123 were false positives (0.4\%), thus showing that non-determinism does not undermine the applicability of \MOTIF.

Though our results may depend on the specific fuzzer used in our experiments, AFL++ is one of the best performing grey-box fuzzers according to recent benchmarks (see Section~\ref{sec:experiment:setup}). Further, though in Section~\ref{sec:symbex} we clarified the technical reasons for not applying hybrid fuzzers, they could be considered in future work if the applicability of their underlying technology (e.g., LLVM) improves.

To address generalizability issues, we selected diverse software subjects that are installed and running on space CPS, including satellites currently in orbit: a mathematical library, a utility library, and a data serialization component. Since they implement a diverse set of features (mathematical operations, serialization, string, and time utilities), they strengthen the generalizability of our results. Further, these types of software components are typical in many CPS systems including avionics, robotics, and automotive, thus suggesting the proposed approach may be useful in many sectors other than space.

\section{Conclusion}
\label{sec:conclusion}

We propose \MOTIF, an approach that leverages fuzzing to automatically generate test data for mutation testing of embedded software deployed in cyber-physical systems (CPS). It aims to overcome the limitations of SOTA approaches, which rely on symbolic execution and cannot easily be applied in many contexts, especially CPS ones. 

\MOTIF is implemented through a pipeline that generates a test driver to process the input data generated by the fuzzer, provides appropriate chunks of such input data 
to the original and mutated versions of a function under test, and determines when the outputs generated by the two functions differ (i.e., the mutant is killed). By monitoring the coverage achieved when executing the original and mutated functions, the fuzzer identifies inputs leading to different behaviors across these functions and, consequently, is driven towards the identification of inputs that kill the mutant.

We performed an empirical evaluation with embedded software deployed on satellites currently in orbit. To compare \MOTIF with a SOTA approach based on symbolic execution, we created an alternative pipeline that leverages symbolic execution instead of fuzzing. Our results show that the approach based on fuzzing outperforms the one based on symbolic execution, for two software subjects where symbolic execution is applicable: it kills 73.79\% and 86.08\% of live mutants in contrast to 26.93\% and 75.56\% for symbolic execution, respectively. Further, it also detects a large number of mutants (35.97\% and 41.38\%) for subjects where symbolic execution  is infeasible. Our results therefore clearly show that fuzzing should be adopted as the preferred method to use to perform mutation testing.  Further, this motivates the development of fuzzing tools dedicated to mutation testing which can, for example, prioritize inputs in the fuzzer queue based on the difference in coverage between the original and the mutated function.

\section*{Acknowledgment}
This research was supported by ESA via a GSTP element contract (RFQ/3-17554/21/NL/AS/kkIMPROVE) and by the NSERC Discovery and Canada Research Chair programs. The authors would like to thank Thierry Titcheu Chekam to help with the development of the SEMUs pipeline.

\bibliographystyle{IEEEtran}
\bibliography{./bibliography/ref}

\end{document}